\documentclass{aa}  

\usepackage{graphicx}
\usepackage{subcaption}
\usepackage{hyperref}
\usepackage{rotating}
\usepackage{xcolor}
\usepackage{amsmath}
\usepackage{CJKutf8}
\usepackage{siunitx}
\usepackage{threeparttable}
\usepackage{comment}
\usepackage{flushend} 

\usepackage{txfonts}

\begin{document} 

   \title{The GECKOS Survey: on the formation history of a barred galaxy via structural decomposition and spatially-resolved spectroscopy}
\titlerunning{GECKOS: PGC\,044931}
\authorrunning{Fraser-McKelvie et al.}

   \author{A. Fraser-McKelvie,\inst{1}   
          D.~A. Gadotti,\inst{2}
          F. Fragkoudi,\inst{3}
          C. de S\'a-Freitas,\inst{4}
          M. Martig,\inst{5}
          M. Bureau, \inst{6}
          T. Davis,\inst{7}
          E. Emsellem, \inst{1}
          R. Elliott,\inst{8}
          D. Fisher,\inst{8}
          M. Hayden,\inst{9}
          J. van de Sande, \inst{10}
          \and A.~B. Watts.\inst{11}
}

\institute{European Southern Observatory, Karl-Schwarzschild-Stra{\ss}e 2, Garching, 85748, Germany\\
    \email{a.fraser-mckelvie@eso.org}  
\and
Centre for Extragalactic Astronomy, Department of Physics, Durham University, South Road, Durham DH1 3LE, UK
\and
Institute for Computational Cosmology, Department of Physics, Durham University, South Road, Durham DH1 3LE, UK 
\and
European Southern Observatory, Alonso de C\'ordova 3107, Casilla 19, Santiago 19001, Chile
\and
Astrophysics Research Institute, Liverpool John Moores University, 146 Brownlow Hill, Liverpool L3 5RF, UK 
\and
Department of Physics, University of Oxford, Denys Wilkinson Building, Keble Road, Oxford OX1 3RH, UK
\and
Cardiff Hub for Astrophysics Research \& Technology, School of Physics \& Astronomy, Cardiff University, Queens Buildings, Cardiff CF24 3AA, UK
\and
Centre for Astrophysics and Supercomputing, Swinburne University of Technology, PO Box 218, Hawthorn, VIC 3122, Australia
\and
Homer L. Dodge Department of Physics \& Astronomy, University of Oklahoma, 440 W.\ Brooks St., Norman, OK 73019, USA
\and 
School of Physics, University of New South Wales, NSW, 2052, Australia
\and
International Centre for Radio Astronomy Research (ICRAR), The University of Western Australia, M468, 35 Stirling Highway, Crawley, WA 6009, Australia\\
             }

  \abstract 
{Disentangling the (co-)evolution of individual galaxy structural components remains a difficult task, owing to the inability to cleanly isolate light from spatially overlapping components. 
In this pilot study of PGC\,044931, observed as part of the GECKOS survey, we utilise a VIRCAM $H$-band image to decompose the galaxy into five photometric components, three of which dominate by contributing $>50\%$ of light in specific regions: a main disc, a boxy/peanut bulge, and a nuclear disc. 
When the photometric decompositions are mapped onto MUSE observations, we find remarkably good separation in stellar kinematic space. All three structures occupy unique locations in the parameter space of the ratio of the light-weighted stellar line-of-sight mean velocity and velocity dispersion ($\rm{V}_{\star}/\sigma_{\star}$), and the high-order stellar skew ($h_{3}$). 
These clear and distinct kinematic behaviours allow us to make inferences about the formation histories of the individual components from observations of the mean stellar ages and metallicities of the three components. 
A clear story emerges: the main disc built over a sustained and extended star formation phase, possibly partly fuelled by gas from a low-metallicity reservoir. Early on, that disc formed a bar that buckled and subsequently formed a nuclear disc in multiple and enriched star-formation episodes.
This result is an example of how careful photometric decompositions, combined with spatially well-resolved stellar kinematic information, can help separate out age-metallicity relations of different components and therefore disentangle the formation history of a galaxy. 
The results of this pilot study can be extended to a differential study of all GECKOS survey galaxies to assert the true diversity of Milky Way-like galaxies.
 }


   \maketitle

\section{Introduction}
An enduring challenge for galaxy evolution studies is extricating the (co)-evolution of individual stellar structures from integrated galaxy light.
Previous statistical photometric and spectroscopic decompositions on 100s of parsec-scale images \citep[e.g.][]{lange2016, haeussler2022} and kiloparsec-scale integral-field spectroscopic (IFS) datasets \citep[e.g.][]{johnston2017, tabor2019, oh2020, pak2021} generally divided galaxies into two components: a (de Vaucouleurs) bulge and an (exponential) disc. These studies were often limited by constraints on signal-to-noise ratio (S/N) and spatial resolution, coupled with a need for automation. 
While statistical in nature, the techniques used often did not account for stellar bars, other non-axisymmetric structures, nor the possibility of a non-dispersion-supported central structure.

Added galactic complexity reveals itself with improved spatial resolution and careful, supervised image decompositions. 
For example, fitting multiple discy components to a galaxy surface brightness profile can often yield a better fit to a galaxy’s light profile than a simple S\'{e}rsic two-component decomposition \citep{gadotti2015}. The era of highly-resolved (10s--100s of parsecs) IFS galaxy surveys has revealed clear signatures of discrete dynamical structures in stellar kinematic maps that can be assigned to features such as nuclear discs\footnote{Here we define a nuclear disc as a central disc (typically built by a bar), of radius 100 pc -- 1 kpc \citep[see review by][]{schultheis2025}.}, stellar bars, boxy/peanut bulges, and axisymmetric discs \citep[e.g.][]{gadotti2020, fraser-mckelvie2025}. Importantly, dynamically-cold central structures such as nuclear discs have been resolved in multiple galaxies \citep[e.g.][though we note that in not all of these instances do we expect these discs to be bar-built]{seidel2015, corsini2016, guerou2016, sarzi2016, pinna2019, martig2021}; and are vital to determine bar ages \citep[e.g.][]{desa-freitas2023, desa-freitas2025}. These results build upon previous theoretical \citep{bureau2005} and observational \citep{chung2004} 1D results linking stellar kinematics along a galaxy’s photometric major axis to stellar structures such as bars and nuclear discs. 

The importance of separating galaxies into their constituent components arises from i) the hypothesis that these structures formed and evolved via different mechanisms, and ii) that some components (e.g.\ stellar bars) may be driving galaxy evolution through internal, secular processes \citep[e.g.][]{athanassoula2013, sellwood2014}.
Therefore, techniques to cleanly separate light and thus isolate meaningful structures, corresponding to different orbital families (remembering that some structures will always be co-spatial regardless of galaxy projection), are paramount.
The result of such will be the ability to infer the star formation histories of individual dynamical stellar structures and understand their formation pathways. 

This paper aims to combine the power of detailed photometric structural decompositions with the information provided by the stellar kinematic and stellar population measurements afforded by IFS observations. In this GECKOS\footnote{GECKOS: Generalising Edge-on galaxies and their Chemical bimodalities, Kinematics, and Outflows out to Solar environments.} pilot study of the boxy/peanut bulge galaxy PGC\,044931, we specifically attempt to understand whether photometric decompositions coupled with IFS measurements can provide meaningful information on the assembly history of individual galaxy structures and attempt to understand the overall galaxy assembly history.

\section{PGC\,044931 - a buckled barred galaxy}
\label{survey_desc}
PGC\,044931 is an edge-on disc galaxy at a distance of 50.7~Mpc \citep{tully2023}. The galaxy possesses a clear boxy/peanut bulge \citep[as seen from unsharp masked imaging in e.g.][]{fraser-mckelvie2025} and is therefore presumed to host a buckled stellar bar \citep{bureau2006}.
\citet{chung2004} confirmed this hypothesis via stellar kinematic measurements of the projected light-weighted line-of-sight mean V$_{\star}$, $\sigma_{\star}$, and high order moment $h_{3}$. They reported a characteristic anti-correlation, correlation, and then anti-correlation of $h_{3}$ vs. $\rm{V}_{\star}/\sigma_{\star}$, as expected for a dynamically-cold nuclear disc \citep[the size of which was measured from 2D IFS observations to be $r=0.71$ kpc;][]{fraser-mckelvie2025}, a stellar bar, and an underlying axisymmetric stellar disc, respectively. The galaxy shows no photometric nor kinematic sign of the presence of a classical bulge. 

\subsection{GECKOS data}
PGC\,044931 was observed with the Very Large Telescope's Multi-Unit Spectroscopic Explorer (MUSE) as part of the GECKOS survey \citep{vandesande2024} via two pointings, the first of which was centred on the bulge, and the second a disc pointing offset to the North-West. 
The raw data were reduced using the ESO MUSE pipeline routines, as packaged by \textsc{Pymusepipe} version 2.23.4 \citep{emsellem2022} and analysed to produce stellar kinematic (V$_{\star}$, $\sigma_{\star}$, $h_{3}$) and mean age and metallicity maps binned to a S/N of 100 using the \textsc{nGIST} pipeline \citep{fraser-mckelvie2025}.  

\subsection{Photometric decomposition}
\begin{figure*}
\centering
\begin{subfigure}{0.89\textwidth}
\includegraphics[width=\textwidth]{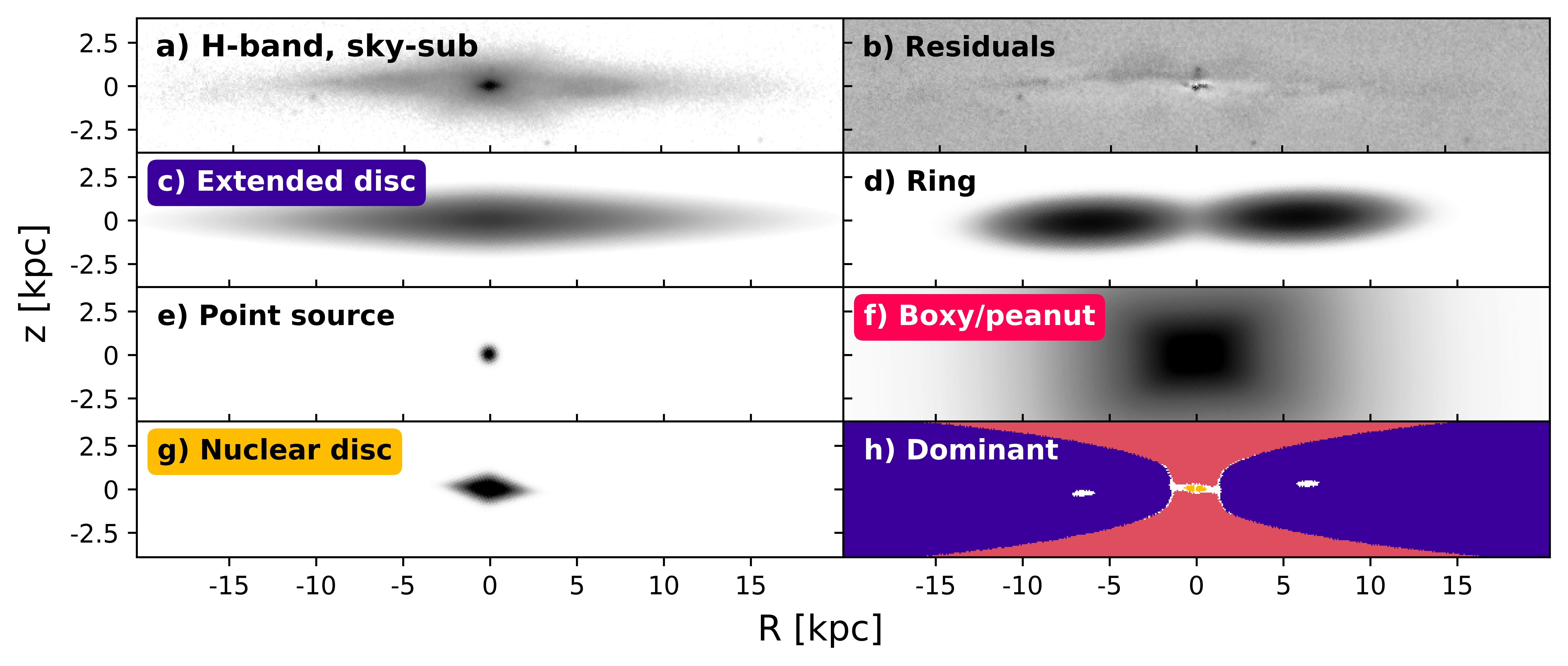}
\end{subfigure}
\caption{Components of the image decomposition of PGC\,044931. Panel a) is the sky-subtracted, $H$-band VIRCAM image (arcsinh scaled), b) is the residual map after subtraction of the five components (linear scale), c)--g) are the five components fit with \textsc{IMFIT} (arcsinh scale, limits altered to show maximum structure), and h) shows the spatial extent of the three dominant components. The extended disc is shown in purple, and the boxy/peanut bulge in pink. The nuclear disc (yellow) is very small and only dominates near the centre. } 
\label{Fig:decomp}
\end{figure*}

Photometric decomposition was performed on an image from the Visible and Infrared Survey Telescope for Astronomy (VISTA), employing the VISTA InfraRed CAMera (VIRCAM). The $H$-band 1.65-$\mu$m image obtained as part of the VISTA hemisphere survey (ESO Program 179.A-2010) with a seeing-limited spatial resolution of 0.8 arcsec (210 parsecs). 
\textsc{IMFIT} \citep{erwin2015a} was employed to model the light profile of the galaxy by adding components until the fit no longer improved. 
The fit was optimised using the Differential Evolution algorithm, which minimises subjective choices, e.g.\ by only requiring lower and upper limits for the fitted parameters from the user (Gadotti 2025, submitted).
The best fit was found to include an extended edge-on disc, an edge-on ring, a central point source, a boxy/peanut bulge, and an edge-on nuclear disc.
A flat sky component that corrects any residuals left from the background subtraction was also included. 

Figure~\ref{Fig:decomp} shows in panel a) the sky-subtracted $H$-band image (arcsinh scaled) and b) the residual image after all components have been subtracted. The 2D light distribution of the five components are shown in panels c)--g), while h) shows the dominant component of each pixel, where the dominant component comprises more than 50\% of the light of that pixel. Only three components ever dominate: the extended disc, the boxy/peanut bulge, and the nuclear disc. White regions correspond to areas in which no component contributes > 50\% of the light and are not considered in the following analysis. We note that the grey contours delineating the spatial regions dominated by the boxy-peanut bulge and extended disc do not directly line up with the `X' shape. We suspect that this discrepancy is due to the 50\% threshold adopted when defining these regions, reflected in the increased metallicity of extended disc data points with a lower dominance fraction seen in Figure~\ref{Fig:scatterplots} (right). Likely there is some contamination from the extended disc in this region.

\begin{figure*}[h]
\centering
\begin{subfigure}{0.95\textwidth}
\includegraphics[width=\textwidth]{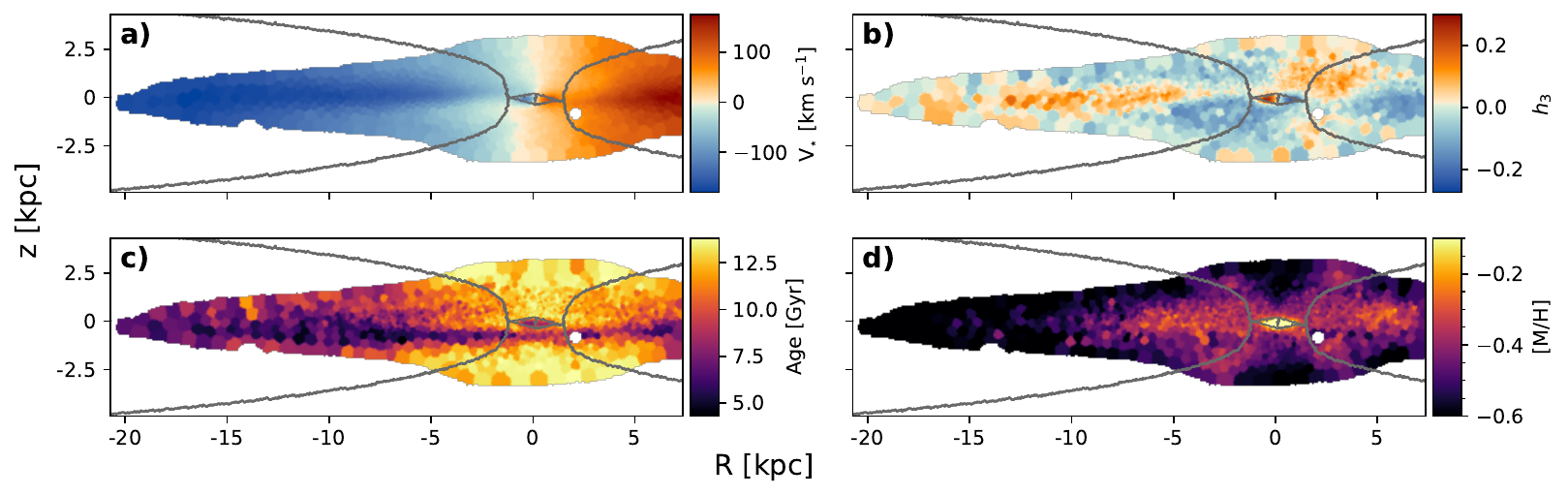}
\end{subfigure}
\caption{Masking regions from Figure~\ref{Fig:decomp}h overlaid as grey contours on light-weighted stellar kinematic (a) V$_{\star}$, (b) $h_{3}$, and population (c) age, and (d) metallicity maps of PGC\,044931. White regions denote masked regions (e.g.\ due to foreground stars), or areas in which the spaxels did not reach the minimum continuum S/N threshold of 5. A distinctive, metal-rich `X' shape is present in the metallicity map, seen in other galaxies \citep[e.g.\ NGC\,4710;][]{gonzalez2016}, and in agreement with simulations \citep[e.g.][]{debattista2017, fragkoudi2020}.} 
\label{Fig:masks}
\end{figure*}

\section{Results \& discussion}
\subsection{Stellar kinematics}

\begin{figure*}
\centering
\begin{subfigure}{0.47\textwidth}
\includegraphics[width=\textwidth]{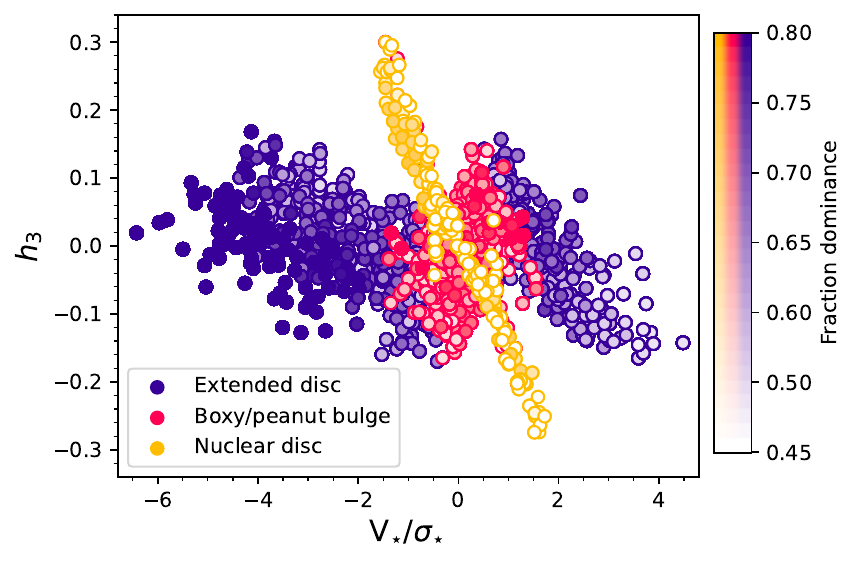}
\end{subfigure}
\begin{subfigure}{0.47\textwidth}
\includegraphics[width=\textwidth]{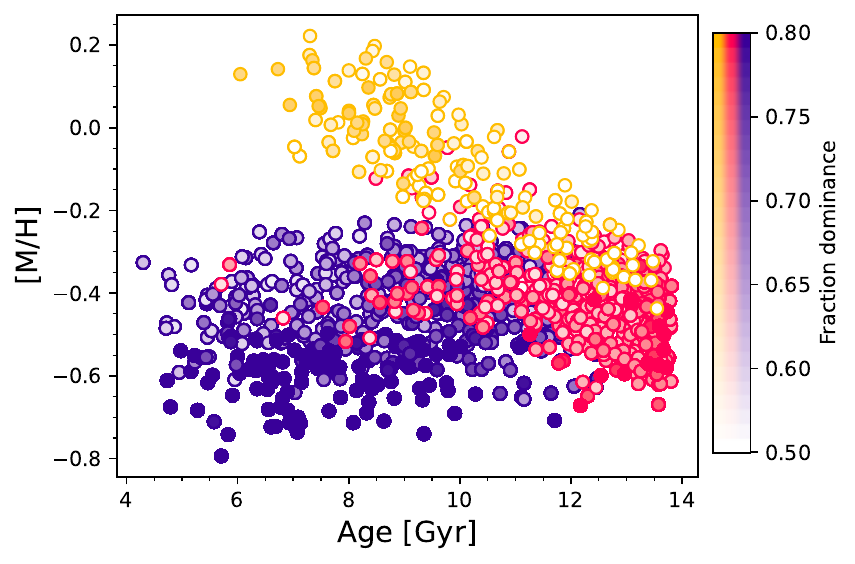}
\end{subfigure}
\caption{Separation of the photometric components of PGC\,044931 in stellar kinematic and stellar population space. Left: stellar $h_{3}$ vs. $\rm{V}_{\star}/\sigma_{\star}$, with data points corresponding to Voronoi bins for which the majority of the bin is located in one of the photometric regions. The underlying extended disc, boxy/peanut bulge (bar), and nuclear disc show distinctly different behaviours in this parameter space, allowing us to confidently make inferences about their star formation histories. Right: mean light-weighted stellar metallicity vs. stellar age. The nuclear disc is clearly separated from the bar and boxy/peanut bulge and the boxy/peanut region is on average older and more metal-rich than the underlying disc. The colour saturation indicates the fraction of light attributed to the given component in each bin.
} 
\label{Fig:scatterplots}
\end{figure*}

We map the dominant component of each VIRCAM pixel onto the Voronoi-binned GECKOS/MUSE stellar kinematic and stellar population maps in a method similar to \citet{querejeta2021} and \citet{martig2021}. 
Maps of light-weighted V$_{\star}$, $h3$, stellar age, and stellar metallicity with the masking regions defined from the photometric decomposition are shown in Figure~\ref{Fig:masks}.
As predicted for non-axisymmetric (bar-like) orbits, the sign of $h_{3}$ matches that of V$_{\star}$ within the boxy-peanut bulge region only.

We then extract the stellar kinematic and population information of each component. In Figure~\ref{Fig:scatterplots} (left panel) we plot $h_{3}$ as a function of $\rm{V}_{\star}/\sigma_{\star}$ for each bin, coloured by the dominant component of each bin, with the opacity of the point representing the fractional dominance of the given component's light in that region. 
All three components exhibit distinct behaviour in this parameter space.   
The nuclear disc (yellow data points) stands out as a clear anti-correlation in $h_{3}-\rm{V}_{\star}/\sigma_{\star}$ for $|\rm{V}_{\star}/\sigma_{\star}|$ $<$2. This behaviour is expected to be the manifestation of the low-velocity tail of the line of sight velocity distribution (LOSVD), attributed to the presence of a decoupled, dynamically cold, rapidly rotating edge-on nuclear disc likely built by the bar \citep{chung2004}. The boxy/peanut bulge (bar) structure (pink data points) shows a correlation between $h_{3}$ and $\rm{V}_{\star}/\sigma_{\star}$, which \citet{bureau2005} attributed to the addition of elongated bar orbits (e.g. x1 and 4:1) to the LOSVD, creating the tail of high-velocity material required for V$_{\star}$ and $h_{3}$ to correlate. 
Finally, as expected, the extended disc shows an anti-correlation between $h_{3}$ and $\rm{V}_{\star}/\sigma_{\star}$ extending out to larger $\rm{V}_{\star}/\sigma_{\star}$, and presumably to larger radii where the disc is more rotation dominated.

Clear separations and distinct behaviours in kinematic parameter space are therefore present between the three photometric components, implying that a simple mapping of the dominant structures extracted from the 2D light distribution can satisfactorily translate to distinct kinematic components.

\subsection{Stellar populations}
We plot the distribution of light-weighted average age and metallicity for each of the three galaxy components (the extended disc, boxy/peanut bulge, and nuclear disc) in Figure~\ref{Fig:scatterplots} (right panel). Again, these three components exhibit distinct behaviours in this parameter space, the division becoming even clearer in regions where a given component has greater dominance. 
While we do not compute uncertainties in this work, similar analyses of MUSE observations report statistical uncertainties of $\approx$1 Gyr in age and 0.05 in [M/H] \citep[e.g.][]{martig2021}.
The extended disc is of similar metallicity to the boxy/peanut bulge, though it spans a range of stellar ages, becoming slightly more metal-poor at younger ages.
The boxy/peanut bulge is almost exclusively old and comparatively metal-poor, and the nuclear disc spans a range of average stellar ages and is more metal-rich than any other region of the galaxy. The divergence of the nuclear disc's age-metallicity relation from the main disc and boxy/peanut bulge occurs at $\approx$10 Gyr. 

Putting the stellar population information together, we can speculate on an evolutionary scenario.
Bars are known to build nuclear discs, which are fed by a protracted supply of cold gas fuelling ongoing star formation as the nuclear disc grows \citep[e.g.][]{hawarden1986, athanassoula1992a, seo2019, verwilghen2024}. 
The measured ages and metallicities require the nuclear disc of PGC\,044931 to have formed more recently than the disc stars that comprise the bar e.g.\ due to star formation in a small physical region that enriches the local area quickly, possibly through processes such as those described in \citet{leaman2019}.
Given the clear offset in metallicity of the nuclear disc from $\approx$10 Gyr, we can infer that the nuclear disc must have formed around that time, and the bar must therefore be at least this old. 

Stars map from the disc to the boxy/peanut bulge at the time of bar buckling, meaning that the stellar population of the boxy/peanut bulge should be representative of the disc at that time \citep[e.g.][]{martinez-valpuesta2013, fragkoudi2017}. To first order, there should be no gas present in the boxy/peanut bulge, hence no further star formation can occur, and the underlying stellar population remains old and metal-poor.  Finally, the extended disc may be the recipient of (near) pristine gas from a broader \ion{H}{i} disc, the cosmic web, or gas-rich minor mergers, allowing ongoing star formation (and hence a range of measured stellar ages) without intensive enrichment of the nuclear disc region. 
Further work to characterise the age of the oldest stellar population in the nuclear disc \citep[and thus infer the age of the bar; e.g.\ ][]{desa-freitas2023, desa-freitas2025} may confirm this scenario. 

\section{Summary \& conclusions}
This Letter presents an example of how careful and detailed photometric decompositions can identify multiple coherent stellar substructures that can be used to understand galaxy assembly history. 

Through the photometric decomposition of a VIRCAM $H$-band image of GECKOS galaxy PGC\,044931, three components were found to dominate the stellar light distribution: an extended disc, a boxy/peanut bulge, and a nuclear disc. 
Mapping the photometric decomposition to GECKOS light-weighted stellar kinematic and stellar age and metallicity maps yielded clear separations between all three components in both kinematic and stellar population parameter space. 
From these decompositions, we propose a scenario in which a disc galaxy formed and and a stellar bar buckled long ago. The bar grew a nuclear disc that enriched via star formation fed by inflowing gas along the bar. The underlying disc exhibits ongoing star formation but little chemical enrichment, likely due to an ongoing supply of (near) pristine gas from an extended \ion{H}{i} disc, the cosmic web, or minor mergers. 

Complementary to recent efforts to combine stellar dynamics and population analyses \citep[e.g.\ ][]{poci2021, ding2023},
this pilot study paves the way for the use of careful photometric decompositions with highly spatially resolved IFS data to understand the assembly history of galactic components. The outcome of which will be a better understanding of the buildup of galactic structures over cosmic time, and the resultant influence of galactic structures on the overall evolution of the host galaxies.

\begin{acknowledgements}
      Based on observations made with ESO telescopes at the La Silla Paranal Observatory under programme IDs 110.24AS and 179.A-2010.
      The authors wish to thank Evelyn Johnston and Boris H\"{a}u{\ss}ler for useful conversations related to the science of this work.
      Author contribution statement: this project was devised by AFM, with analysis completed by AFM, photometric decomposition performed by DAG, and science comments by the whole author list. 
    AFM gratefully acknowledges the sponsorship provided by the European Southern Observatory through a research fellowship. 
    DAG is supported by STFC grant ST/X001075/1.
    FF is supported by a UKRI Future Leaders Fellowship (grant no. MR/X033740/1).

\end{acknowledgements}

\bibliographystyle{aa}
\bibliography{references_pgc}

@article{athanassoula1992a,
	title = {Morphology of bar orbits.},
	volume = {259},
	issn = {0035-8711},
	url = {https://ui.adsabs.harvard.edu/abs/1992MNRAS.259..328A},
	doi = {10.1093/mnras/259.2.328},
	urldate = {2024-05-20},
	journal = {\mnras},
	author = {Athanassoula, E.},
	month = nov,
	year = {1992},
	pages = {328--344},
}

@INCOLLECTION{athanassoula2013,
       author = {{Athanassoula}, E.},
        title = "{Bars and secular evolution in disk galaxies: Theoretical input}",
    booktitle = {Secular Evolution of Galaxies},
         year = 2013,
       editor = {{Falc{\'o}n-Barroso}, Jes{\'u}s and {Knapen}, Johan H.},
        pages = {305},
          doi = {10.48550/arXiv.1211.6752},
       adsurl = {https://ui.adsabs.harvard.edu/abs/2013seg..book..305A}
}

@article{bureau2005,
	title = {Bar {Diagnostics} in {Edge}-{On} {Spiral} {Galaxies}. {III}. {N}-{Body} {Simulations} of {Disks}},
	volume = {626},
	issn = {0004-637X},
	url = {https://ui.adsabs.harvard.edu/abs/2005ApJ...626..159B},
	doi = {10.1086/430056},
	urldate = {2023-10-05},
	journal = {\apj},
	author = {Bureau, M. and Athanassoula, E.},
	month = jun,
	year = {2005},
	pages = {159--173},
}

@ARTICLE{bureau2006,
       author = {{Bureau}, M. and {Aronica}, G. and {Athanassoula}, E. and {Dettmar}, R. -J. and {Bosma}, A. and {Freeman}, K.~C.},
        title = "{K-band observations of boxy bulges - I. Morphology and surface brightness profiles}",
      journal = {\mnras},
         year = 2006,
        month = aug,
       volume = {370},
       number = {2},
        pages = {753-772},
          doi = {10.1111/j.1365-2966.2006.10471.x},
archivePrefix = {arXiv},
       eprint = {astro-ph/0606056},
 primaryClass = {astro-ph},
       adsurl = {https://ui.adsabs.harvard.edu/abs/2006MNRAS.370..753B}
}

@article{chung2004,
	title = {Stellar {Kinematics} of {Boxy} {Bulges}: {Large}-{Scale} {Bars} and {Inner} {Disks}},
	volume = {127},
	issn = {0004-6256},
	shorttitle = {Stellar {Kinematics} of {Boxy} {Bulges}},
	url = {https://ui.adsabs.harvard.edu/abs/2004AJ....127.3192C},
	doi = {10.1086/420988},
	urldate = {2024-01-12},
	journal = {\aj},
	author = {Chung, Aeree and Bureau, M.},
	month = jun,
	year = {2004},
	pages = {3192--3212},
}

@ARTICLE{corsini2016,
       author = {{Corsini}, E.~M. and {Morelli}, L. and {Pastorello}, N. and {Dalla Bont{\`a}}, E. and {Pizzella}, A. and {Portaluri}, E.},
        title = "{The young nuclear stellar disc in the SB0 galaxy NGC 1023}",
      journal = {\mnras},
         year = 2016,
        month = apr,
       volume = {457},
       number = {2},
        pages = {1198-1207},
          doi = {10.1093/mnras/stv2864},
archivePrefix = {arXiv},
       eprint = {1511.08219},
 primaryClass = {astro-ph.GA},
       adsurl = {https://ui.adsabs.harvard.edu/abs/2016MNRAS.457.1198C}
}

@ARTICLE{debattista2017,
       author = {{Debattista}, Victor P. and {Ness}, Melissa and {Gonzalez}, Oscar A. and {Freeman}, K. and {Zoccali}, Manuela and {Minniti}, Dante},
        title = "{Separation of stellar populations by an evolving bar: implications for the bulge of the Milky Way}",
      journal = {\mnras},
         year = 2017,
        month = aug,
       volume = {469},
       number = {2},
        pages = {1587-1611},
          doi = {10.1093/mnras/stx947},
archivePrefix = {arXiv},
       eprint = {1611.09023},
 primaryClass = {astro-ph.GA},
       adsurl = {https://ui.adsabs.harvard.edu/abs/2017MNRAS.469.1587D}
}

@article{desa-freitas2023,
	title = {Disc galaxies are still settling. {Discovery} of the smallest nuclear discs and their young stellar bars},
	volume = {678},
	issn = {0004-6361},
	url = {https://ui.adsabs.harvard.edu/abs/2023A&A...678A.202D},
	doi = {10.1051/0004-6361/202347028},
	urldate = {2024-10-29},
	journal = {\aap},
	author = {de Sá-Freitas, Camila and Gadotti, Dimitri A. and Fragkoudi, Francesca and Coccato, Lodovico and Coelho, Paula and de Lorenzo-Cáceres, Adriana and Falcón-Barroso, Jesús and Kolcu, Tutku and Martín-Navarro, Ignacio and Mendez-Abreu, Jairo and Neumann, Justus and Blazquez, Patricia Sanchez and Querejeta, Miguel and van de Ven, Glenn},
	month = oct,
	year = {2023},
	pages = {A202},
}

@article{desa-freitas2025,
	title = {Bar ages derived for the first time in nearby galaxies: {Insights} into secular evolution from the {TIMER} sample},
	volume = {698},
	issn = {0004-6361},
	shorttitle = {Bar ages derived for the first time in nearby galaxies},
	url = {https://ui.adsabs.harvard.edu/abs/2025A&A...698A...5D},
	doi = {10.1051/0004-6361/202453367},
	urldate = {2025-08-27},
	journal = {\aap},
	author = {de Sá-Freitas, Camila and Gadotti, Dimitri A. and Fragkoudi, Francesca and Coelho, Paula and de Lorenzo-Cáceres, Adriana and Falcón-Barroso, Jesús and Sánchez-Blázquez, Patricia and Kim, Taehyun and Mendez-Abreu, Jairo and Neumann, Justus and Querejeta, Miguel and van de Ven, Glenn},
	month = jun,
	year = {2025},
	pages = {A5},
}

@ARTICLE{ding2023,
       author = {{Ding}, Y. and {Zhu}, L. and {van de Ven}, G. and {Coccato}, L. and {Corsini}, E.~M. and {Costantin}, L. and {Fahrion}, K. and {Falc{\'o}n-Barroso}, J. and {Gadotti}, D.~A. and {Iodice}, E. and {Lyubenova}, M. and {Mart{\'\i}n-Navarro}, I. and {McDermid}, R.~M. and {Pinna}, F. and {Sarzi}, M.},
        title = "{The Fornax3D project: Environmental effects on the assembly of dynamically cold disks in Fornax cluster galaxies}",
      journal = {\aap},
         year = 2023,
        month = apr,
       volume = {672},
          eid = {A84},
        pages = {A84},
          doi = {10.1051/0004-6361/202244558},
archivePrefix = {arXiv},
       eprint = {2301.05532},
 primaryClass = {astro-ph.GA},
       adsurl = {https://ui.adsabs.harvard.edu/abs/2023A&A...672A..84D}
}

@article{emsellem2022,
	title = {The {PHANGS}-{MUSE} survey. {Probing} the chemo-dynamical evolution of disc galaxies},
	volume = {659},
	issn = {0004-6361},
	url = {https://ui.adsabs.harvard.edu/abs/2022A&A...659A.191E},
	doi = {10.1051/0004-6361/202141727},
	urldate = {2023-09-25},
	journal = {\aap},
	author = {Emsellem, Eric and Schinnerer, Eva and Santoro, Francesco and Belfiore, Francesco and Pessa, Ismael and McElroy, Rebecca and Blanc, Guillermo A. and Congiu, Enrico and Groves, Brent and Ho, I. -Ting and Kreckel, Kathryn and Razza, Alessandro and Sanchez-Blazquez, Patricia and Egorov, Oleg and Faesi, Chris and Klessen, Ralf S. and Leroy, Adam K. and Meidt, Sharon and Querejeta, Miguel and Rosolowsky, Erik and Scheuermann, Fabian and Anand, Gagandeep S. and Barnes, Ashley T. and Bešlić, Ivana and Bigiel, Frank and Boquien, Médéric and Cao, Yixian and Chevance, Mélanie and Dale, Daniel A. and Eibensteiner, Cosima and Glover, Simon C. O. and Grasha, Kathryn and Henshaw, Jonathan D. and Hughes, Annie and Koch, Eric W. and Kruijssen, J. M. Diederik and Lee, Janice and Liu, Daizhong and Pan, Hsi-An and Pety, Jérôme and Saito, Toshiki and Sandstrom, Karin M. and Schruba, Andreas and Sun, Jiayi and Thilker, David A. and Usero, Antonio and Watkins, Elizabeth J. and Williams, Thomas G.},
	month = mar,
	year = {2022},
	pages = {A191},
}

@article{erwin2015a,
	title = {Composite bulges: the coexistence of classical bulges and discy pseudo-bulges in {S0} and spiral galaxies},
	volume = {446},
	issn = {0035-8711},
	shorttitle = {Composite bulges},
	url = {https://ui.adsabs.harvard.edu/abs/2015MNRAS.446.4039E},
	doi = {10.1093/mnras/stu2376},
	urldate = {2024-08-19},
	journal = {\mnras},
	author = {Erwin, Peter and Saglia, Roberto P. and Fabricius, Maximilian and Thomas, Jens and Nowak, Nina and Rusli, Stephanie and Bender, Ralf and Vega Beltrán, Juan Carlos and Beckman, John E.},
	month = feb,
	year = {2015},
	pages = {4039--4077},
}

@article{fragkoudi2017,
	title = {What the {Milky} {Way} bulge reveals about the initial metallicity gradients in the disc},
	volume = {607},
	issn = {0004-6361},
	url = {https://ui.adsabs.harvard.edu/abs/2017A&A...607L...4F},
	doi = {10.1051/0004-6361/201731597},
	urldate = {2023-09-29},
	journal = {\aap},
	author = {Fragkoudi, F. and Di Matteo, P. and Haywood, M. and Khoperskov, S. and Gomez, A. and Schultheis, M. and Combes, F. and Semelin, B.},
	month = nov,
	year = {2017},
	pages = {L4},
}

@article{fragkoudi2020,
	title = {Chemodynamics of barred galaxies in cosmological simulations: {On} the {Milky} {Way}'s quiescent merger history and in-situ bulge},
	volume = {494},
	issn = {0035-8711},
	shorttitle = {Chemodynamics of barred galaxies in cosmological simulations},
	url = {https://ui.adsabs.harvard.edu/abs/2020MNRAS.494.5936F},
	doi = {10.1093/mnras/staa1104},
	urldate = {2022-03-11},
	journal = {\mnras},
	author = {Fragkoudi, F. and Grand, R. J. J. and Pakmor, R. and Blázquez-Calero, G. and Gargiulo, I. and Gomez, F. and Marinacci, F. and Monachesi, A. and Ness, M. K. and Perez, I. and Tissera, P. and White, S. D. M.},
	month = jun,
	year = {2020},
	pages = {5936--5960},
}

@ARTICLE{fraser-mckelvie2025,
       author = {{Fraser-McKelvie}, A. and {van de Sande}, J. and {Gadotti}, D.~A. and {Emsellem}, E. and {Brown}, T. and {Fisher}, D.~B. and {Martig}, M. and {Bureau}, M. and {Gerhard}, O. and {Battisti}, A.~J. and {Bland-Hawthorn}, J. and {Boecker}, A. and {Catinella}, B. and {Combes}, F. and {Cortese}, L. and {Croom}, S.~M. and {Davis}, T.~A. and {Falc{\'o}n-Barroso}, J. and {Fragkoudi}, F. and {Freeman}, K.~C. and {Hayden}, M.~R. and {McDermid}, R. and {Mazzilli Ciraulo}, B. and {Mendel}, J.~T. and {Pinna}, F. and {Poci}, A. and {Rutherford}, T.~H. and {de S{\'a}-Freitas}, C. and {Silva-Lima}, L.~A. and {Valenzuela}, L.~M. and {van de Ven}, G. and {Wang}, Z. and {Watts}, A.~B.},
        title = "{The GECKOS survey: Identifying kinematic sub-structures in edge-on galaxies}",
      journal = {\aap},
         year = 2025,
        month = aug,
       volume = {700},
          eid = {A237},
        pages = {A237},
          doi = {10.1051/0004-6361/202452891},
archivePrefix = {arXiv},
       eprint = {2411.03430},
 primaryClass = {astro-ph.GA},
       adsurl = {https://ui.adsabs.harvard.edu/abs/2025A&A...700A.237F}
}

@article{gadotti2020,
	title = {Kinematic signatures of nuclear discs and bar-driven secular evolution in nearby galaxies of the {MUSE} {TIMER} project},
	volume = {643},
	issn = {0004-6361},
	url = {https://ui.adsabs.harvard.edu/abs/2020A&A...643A..14G},
	doi = {10.1051/0004-6361/202038448},
	urldate = {2023-12-19},
	journal = {\aap},
	author = {Gadotti, Dimitri A. and Bittner, Adrian and Falcón-Barroso, Jesús and Méndez-Abreu, Jairo and Kim, Taehyun and Fragkoudi, Francesca and de Lorenzo-Cáceres, Adriana and Leaman, Ryan and Neumann, Justus and Querejeta, Miguel and Sánchez-Blázquez, Patricia and Martig, Marie and Martín-Navarro, Ignacio and Pérez, Isabel and Seidel, Marja K. and van de Ven, Glenn},
	month = nov,
	year = {2020},
	pages = {A14},
}

@ARTICLE{gadotti2015,
       author = {{Gadotti}, Dimitri A. and {Seidel}, Marja K. and {S{\'a}nchez-Bl{\'a}zquez}, Patricia and {Falc{\'o}n-Barroso}, Jesus and {Husemann}, Bernd and {Coelho}, Paula and {P{\'e}rez}, Isabel},
        title = "{MUSE tells the story of NGC 4371: The dawning of secular evolution}",
      journal = {\aap},
         year = 2015,
        month = dec,
       volume = {584},
          eid = {A90},
        pages = {A90},
          doi = {10.1051/0004-6361/201526677},
archivePrefix = {arXiv},
       eprint = {1509.00032},
 primaryClass = {astro-ph.GA},
       adsurl = {https://ui.adsabs.harvard.edu/abs/2015A&A...584A..90G}
}

@ARTICLE{gonzalez2016,
       author = {{Gonzalez}, O.~A. and {Gadotti}, D.~A. and {Debattista}, V.~P. and {Rejkuba}, M. and {Valenti}, E. and {Zoccali}, M. and {Coccato}, L. and {Minniti}, D. and {Ness}, M.},
        title = "{Comparing the properties of the X-shaped bulges of NGC 4710 and the Milky Way with MUSE}",
      journal = {\aap},
         year = 2016,
        month = jun,
       volume = {591},
          eid = {A7},
        pages = {A7},
          doi = {10.1051/0004-6361/201527806},
archivePrefix = {arXiv},
       eprint = {1603.02546},
 primaryClass = {astro-ph.GA},
       adsurl = {https://ui.adsabs.harvard.edu/abs/2016A&A...591A...7G}
}

@ARTICLE{hawarden1986,
       author = {{Hawarden}, T.~G. and {Mountain}, C.~M. and {Leggett}, S.~K. and {Puxley}, P.~J.},
        title = "{Enhanced star formation - the importance of bars in spiral galaxies.}",
      journal = {\mnras},
         year = 1986,
        month = aug,
       volume = {221},
        pages = {41P-45},
          doi = {10.1093/mnras/221.1.41P},
       adsurl = {https://ui.adsabs.harvard.edu/abs/1986MNRAS.221P..41H}
}

@ARTICLE{guerou2016,
       author = {{Gu{\'e}rou}, A. and {Emsellem}, E. and {Krajnovi{\'c}}, D. and {McDermid}, R.~M. and {Contini}, T. and {Weilbacher}, P.~M.},
        title = "{Exploring the mass assembly of the early-type disc galaxy NGC 3115 with MUSE}",
      journal = {\aap},
         year = 2016,
        month = jul,
       volume = {591},
          eid = {A143},
        pages = {A143},
          doi = {10.1051/0004-6361/201628743},
archivePrefix = {arXiv},
       eprint = {1605.07667},
 primaryClass = {astro-ph.GA},
       adsurl = {https://ui.adsabs.harvard.edu/abs/2016A&A...591A.143G}
}

@ARTICLE{haeussler2022,
       author = {{H{\"a}u{\ss}ler}, Boris and {Vika}, Marina and {Bamford}, Steven P. and {Johnston}, Evelyn J. and {Brough}, Sarah and {Casura}, Sarah and {Holwerda}, Benne W. and {Kelvin}, Lee S. and {Popescu}, Cristina},
        title = "{GALAPAGOS-2/GALFITM/GAMA - Multi-wavelength measurement of galaxy structure: Separating the properties of spheroid and disk components in modern surveys}",
      journal = {\aap},
         year = 2022,
        month = aug,
       volume = {664},
          eid = {A92},
        pages = {A92},
          doi = {10.1051/0004-6361/202142935},
archivePrefix = {arXiv},
       eprint = {2204.05907},
 primaryClass = {astro-ph.GA},
       adsurl = {https://ui.adsabs.harvard.edu/abs/2022A&A...664A..92H}
}

@ARTICLE{johnston2017,
       author = {{Johnston}, Evelyn J. and {H{\"a}u{\ss}ler}, Boris and {Arag{\'o}n-Salamanca}, Alfonso and {Merrifield}, Michael R. and {Bamford}, Steven and {Bershady}, Matthew A. and {Bundy}, Kevin and {Drory}, Niv and {Fu}, Hai and {Law}, David and {Nitschelm}, Christian and {Thomas}, Daniel and {Roman Lopes}, Alexandre and {Wake}, David and {Yan}, Renbin},
        title = "{SDSS-IV MaNGA: bulge-disc decomposition of IFU data cubes (BUDDI)}",
      journal = {\mnras},
         year = 2017,
        month = feb,
       volume = {465},
       number = {2},
        pages = {2317-2341},
          doi = {10.1093/mnras/stw2823},
archivePrefix = {arXiv},
       eprint = {1611.00609},
 primaryClass = {astro-ph.GA},
       adsurl = {https://ui.adsabs.harvard.edu/abs/2017MNRAS.465.2317J}
}

@ARTICLE{lange2016,
       author = {{Lange}, Rebecca and {Moffett}, Amanda J. and {Driver}, Simon P. and {Robotham}, Aaron S.~G. and {Lagos}, Claudia del P. and {Kelvin}, Lee S. and {Conselice}, Christopher and {Margalef-Bentabol}, Berta and {Alpaslan}, Mehmet and {Baldry}, Ivan and {Bland-Hawthorn}, Joss and {Bremer}, Malcolm and {Brough}, Sarah and {Cluver}, Michelle and {Colless}, Matthew and {Davies}, Luke J.~M. and {H{\"a}u{\ss}ler}, Boris and {Holwerda}, Benne W. and {Hopkins}, Andrew M. and {Kafle}, Prajwal R. and {Kennedy}, Rebecca and {Liske}, Jochen and {Phillipps}, Steven and {Popescu}, Cristina C. and {Taylor}, Edward N. and {Tuffs}, Richard and {van Kampen}, Eelco and {Wright}, Angus H.},
        title = "{Galaxy And Mass Assembly (GAMA): M\_star - R\_e relations of z = 0 bulges, discs and spheroids}",
      journal = {\mnras},
         year = 2016,
        month = oct,
       volume = {462},
       number = {2},
        pages = {1470-1500},
          doi = {10.1093/mnras/stw1495},
archivePrefix = {arXiv},
       eprint = {1607.01096},
 primaryClass = {astro-ph.GA},
       adsurl = {https://ui.adsabs.harvard.edu/abs/2016MNRAS.462.1470L}
}

@ARTICLE{leaman2019,
       author = {{Leaman}, Ryan and {Fragkoudi}, Francesca and {Querejeta}, Miguel and {Leung}, Gigi Y.~C. and {Gadotti}, Dimitri A. and {Husemann}, Bernd and {Falc{\'o}n-Barroso}, Jesus and {S{\'a}nchez-Bl{\'a}zquez}, Patricia and {van de Ven}, Glenn and {Kim}, Taehyun and {Coelho}, Paula and {Lyubenova}, Mariya and {de Lorenzo-C{\'a}ceres}, Adriana and {Martig}, Marie and {Martinez-Valpuesta}, Inma and {Neumann}, Justus and {P{\'e}rez}, Isabel and {Seidel}, Marja},
        title = "{Survival of molecular gas in a stellar feedback-driven outflow witnessed with the MUSE TIMER project and ALMA}",
      journal = {\mnras},
         year = 2019,
        month = sep,
       volume = {488},
       number = {3},
        pages = {3904-3928},
          doi = {10.1093/mnras/stz1844},
archivePrefix = {arXiv},
       eprint = {1907.13142},
 primaryClass = {astro-ph.GA},
       adsurl = {https://ui.adsabs.harvard.edu/abs/2019MNRAS.488.3904L}
}

@article{martig2021,
	title = {{NGC} 5746: {Formation} history of a massive disc-dominated galaxy},
	volume = {508},
	issn = {0035-8711},
	shorttitle = {{NGC} 5746},
	url = {https://ui.adsabs.harvard.edu/abs/2021MNRAS.508.2458M},
	doi = {10.1093/mnras/stab2729},
	urldate = {2023-09-25},
	journal = {\mnras},
	author = {Martig, Marie and Pinna, Francesca and Falcón-Barroso, Jesús and Gadotti, Dimitri A. and Husemann, Bernd and Minchev, Ivan and Neumann, Justus and Ruiz-Lara, Tomás and van de Ven, Glenn},
	month = dec,
	year = {2021},
	pages = {2458--2478},
}

@article{martinez-valpuesta2013,
	title = {Metallicity {Gradients} {Through} {Disk} {Instability}: {A} {Simple} {Model} for the {Milky} {Way}'s {Boxy} {Bulge}},
	volume = {766},
	issn = {0004-637X},
	shorttitle = {Metallicity {Gradients} {Through} {Disk} {Instability}},
	url = {https://ui.adsabs.harvard.edu/abs/2013ApJ...766L...3M},
	doi = {10.1088/2041-8205/766/1/L3},
	urldate = {2023-10-01},
	journal = {\apj},
	author = {Martinez-Valpuesta, Inma and Gerhard, Ortwin},
	month = mar,
	year = {2013},
	pages = {L3},
}

@article{oh2020,
	title = {The {SAMI} {Galaxy} {Survey}: decomposed stellar kinematics of galaxy bulges and disks},
	volume = {495},
	issn = {0035-8711},
	shorttitle = {The {SAMI} {Galaxy} {Survey}},
	url = {http://adsabs.harvard.edu/abs/2020MNRAS.495.4638O},
	doi = {10.1093/mnras/staa1330},
	urldate = {2020-12-22},
	journal = {\mnras},
	author = {Oh, Sree and Colless, Matthew and Barsanti, Stefania and Casura, Sarah and Cortese, Luca and van de Sande, Jesse and Owers, Matt S. and Scott, Nicholas and D'Eugenio, Francesco and Bland-Hawthorn, Joss and Brough, Sarah and Bryant, Julia J. and Croom, Scott M. and Foster, Caroline and Groves, Brent and Lawrence, Jon S. and Richards, Samuel N. and Sweet, Sarah M.},
	month = may,
	year = {2020},
	pages = {4638--4658},
}

@ARTICLE{pak2021,
       author = {{Pak}, Mina and {Lee}, Joon Hyeop and {Oh}, Sree and {D'Eugenio}, Francesco and {Colless}, Matthew and {Jeong}, Hyunjin and {Jeong}, Woong-Seob},
        title = "{Stellar Populations of Spectroscopically Decomposed Bulge-Disk for S0 Galaxies from the CALIFA Survey}",
      journal = {\apj},
         year = 2021,
        month = nov,
       volume = {921},
       number = {1},
          eid = {49},
        pages = {49},
          doi = {10.3847/1538-4357/ac1ba1},
archivePrefix = {arXiv},
       eprint = {2108.05014},
 primaryClass = {astro-ph.GA},
       adsurl = {https://ui.adsabs.harvard.edu/abs/2021ApJ...921...49P}
}

@ARTICLE{pinna2019,
       author = {{Pinna}, F. and {Falc{\'o}n-Barroso}, J. and {Martig}, M. and {Sarzi}, M. and {Coccato}, L. and {Iodice}, E. and {Corsini}, E.~M. and {de Zeeuw}, P.~T. and {Gadotti}, D.~A. and {Leaman}, R. and {Lyubenova}, M. and {McDermid}, R.~M. and {Minchev}, I. and {Morelli}, L. and {van de Ven}, G. and {Viaene}, S.},
        title = "{The Fornax 3D project: Unveiling the thick disk origin in FCC 170; possible signs of accretion}",
      journal = {\aap},
         year = 2019,
        month = mar,
       volume = {623},
          eid = {A19},
        pages = {A19},
          doi = {10.1051/0004-6361/201833193},
archivePrefix = {arXiv},
       eprint = {1901.04310},
 primaryClass = {astro-ph.GA},
       adsurl = {https://ui.adsabs.harvard.edu/abs/2019A&A...623A..19P}
}

@ARTICLE{poci2021,
       author = {{Poci}, A. and {McDermid}, R.~M. and {Lyubenova}, M. and {Zhu}, L. and {van de Ven}, G. and {Iodice}, E. and {Coccato}, L. and {Pinna}, F. and {Corsini}, E.~M. and {Falc{\'o}n-Barroso}, J. and {Gadotti}, D.~A. and {Grand}, R.~J.~J. and {Fahrion}, K. and {Mart{\'\i}n-Navarro}, I. and {Sarzi}, M. and {Viaene}, S. and {de Zeeuw}, P.~T.},
        title = "{The Fornax3D project: Assembly histories of lenticular galaxies from a combined dynamical and population orbital analysis}",
      journal = {\aap},
         year = 2021,
        month = mar,
       volume = {647},
          eid = {A145},
        pages = {A145},
          doi = {10.1051/0004-6361/202039644},
archivePrefix = {arXiv},
       eprint = {2102.02449},
 primaryClass = {astro-ph.GA},
       adsurl = {https://ui.adsabs.harvard.edu/abs/2021A&A...647A.145P}
}

@ARTICLE{sarzi2016,
       author = {{Sarzi}, M. and {Ledo}, H.~R. and {Coccato}, L. and {Corsini}, E.~M. and {Dotti}, M. and {Khochfar}, S. and {Maraston}, C. and {Morelli}, L. and {Pizzella}, A.},
        title = "{Nuclear discs as clocks for the assembly history of early-type galaxies: the case of NGC 4458}",
      journal = {\mnras},
         year = 2016,
        month = apr,
       volume = {457},
       number = {2},
        pages = {1804-1812},
          doi = {10.1093/mnras/stw099},
archivePrefix = {arXiv},
       eprint = {1601.03292},
 primaryClass = {astro-ph.GA},
       adsurl = {https://ui.adsabs.harvard.edu/abs/2016MNRAS.457.1804S}
}

\end{document}